\documentclass[aps,pccp,preprint,superscriptaddress]{revtex4-1}
\usepackage{graphicx} 
\usepackage{amsmath}
\usepackage[version=3]{mhchem}
\usepackage{longtable}
\usepackage{filecontents}
\usepackage{array}
\usepackage[referable]{threeparttablex}
\usepackage{changepage}
\usepackage[left=1.5cm, right=1.5cm, top=2.4cm, bottom=2.4cm]{geometry}

\begin{document}

\title{Assessment of density functional methods with correct asymptotic behavior} 

\author{Chen-Wei Tsai} 
\affiliation{Department of Physics, National Taiwan University, Taipei 10617, Taiwan} 

\author{Yu-Chuan Su} 
\affiliation{Department of Physics, National Taiwan University, Taipei 10617, Taiwan} 

\author{Guan-De Li} 
\affiliation{Department of Physics, National Taiwan University, Taipei 10617, Taiwan} 

\author{Jeng-Da Chai} 
\email[Author to whom correspondence should be addressed. Electronic mail: ]{jdchai@phys.ntu.edu.tw.} 
\affiliation{Department of Physics, National Taiwan University, Taipei 10617, Taiwan} 
\affiliation{Center for Theoretical Sciences and Center for Quantum Science and Engineering, National Taiwan University, Taipei 10617, Taiwan} 

\date{\today}

\begin{abstract}

Long-range corrected (LC) hybrid functionals and asymptotically corrected (AC) model potentials are two distinct density functional methods with correct asymptotic behavior. They are known to be accurate 
for properties that are sensitive to the asymptote of the exchange-correlation potential, such as the highest occupied molecular orbital energies and Rydberg excitation energies of molecules. 
To provide a comprehensive comparison, we investigate the performance of the two schemes and others on a very wide range of applications, including the asymptote problems, 
self-interaction-error problems, energy-gap problems, charge-transfer problems, and many others. The LC hybrid scheme is shown to consistently outperform the AC model potential scheme. In addition, 
to be consistent with the molecules collected in the IP131 database [Y.-S. Lin, C.-W. Tsai, G.-D. Li, and J.-D. Chai, \textit{J. Chem. Phys.}, 2012, {\bf 136}, 154109], we expand the EA115 and FG115 databases to include, 
respectively, the vertical electron affinities and fundamental gaps of the additional 16 molecules, and develop a new database AE113 (113 atomization energies), consisting of accurate reference values for 
the atomization energies of the 113 molecules in IP131. These databases will be useful for assessing the accuracy of density functional methods. 

\end{abstract}

\maketitle

\section{Introduction}

Over the past two decades, Kohn-Sham density functional theory (KS-DFT) \cite{HK,KS} has been one of the most powerful approaches for the study of large ground-state systems due to its low computational cost and 
reasonable accuracy \cite{Parr,DFTReview,Perdew09}. Its extension to the real-time domain, time-dependent density functional theory (TDDFT) \cite{TDDFT}, has also been actively developed for studying 
the time-dependent and excited-state properties of large systems \cite{TDDFT-2,TDDFT-3,Casida,Gross}. 

In KS-DFT, the exact exchange-correlation (XC) energy functional $E_{xc}[\rho]$ remains unknown and needs to be approximated. Accurate approximations to $E_{xc}[\rho]$ have been successively developed 
to extend the applicability of KS-DFT to a wide range of systems. Among them, semilocal density functionals \cite{semilocal} perform satisfactorily for some applications but can produce erroneous results in situations 
where the accurate treatment of nonlocality of the XC hole is important \cite{SciYang,DFTreview2,Chai2012}. 

One of the most important and challenging topics in KS-DFT is the asymptotic behavior of the XC potential $v_{xc}(\textbf{r}) = \delta E_{xc}[\rho]/\delta \rho(\textbf{r})$. For finite systems, the exact $v_{xc}(\textbf{r})$ 
should exhibit the Coulombic $(-1/r)$ decay as $r \rightarrow \infty$ \cite{exactip,asymp,sum_vxc2,sum_vxc3}. However, owing to the severe self-interaction error (SIE) \cite{SIC}, the XC potential of semilocal functionals 
decays much faster than $(-1/r)$ in the asymptotic region, yielding qualitatively incorrect predictions for properties sensitive to the asymptote of the XC potential, such as the highest occupied molecular orbital (HOMO) 
energies and high-lying (Rydberg) excitation energies of atoms and molecules \cite{HOMO_BAD,failure_LG}. 

To date, perhaps the most successful density functional methods in practice to include correct XC potential in the asymptotic region of molecules are provided by the long-range corrected (LC) hybrid 
scheme \cite{LC-DFT,LCHirao,CAM-B3LYP,LC-wPBE,BNL,wB97X,wB97X-D,op,wB97X-2,wM05-D,LC-D3} and asymptotically corrected (AC) model potential scheme \cite{LB94,LBa,BJ,AC,AC3,AC4,AC6,AC7,AC8,AC9}. 
For the LC hybrid scheme, the nonlocal Hartree-Fock (HF) exchange for the long-range electron-electron interactions is added to a semilocal functional, which thereby resolves a significant part of the SIE problems 
and provides an AC XC potential. By contrast, for the AC model potential scheme, an AC XC potential is directly modeled and added to a semilocal functional. In principle, a model XC potential should be a 
functional derivative of some $E_{xc}[\rho]$. However, several existing AC model potentials are found {\it not} to be functional derivatives. Consequently, when these AC model potentials are employed, 
several necessary conditions for a functional derivative, such as (i) minimization of the ground-state energy by the self-consistent KS orbitals, (ii) path independence of the van Leeuwen-Baerends line integral \cite{LB95}, 
and (iii) net zero force and zero torque on the ground-state density, can be violated \cite{Staroverov}. Besides, as these model potentials are not variationally stable, the associated XC energies are not well-defined, 
and the predicted properties need to be carefully interpreted. The proposals made by Staroverov and co-workers \cite{Staroverov2a,Staroverov2b} are useful for transforming a model GGA exchange potential (which is not 
a functional derivative) into an exchange potential that has a parent GGA functional. However, for the LB94 potential (an AC model potential with no parent GGA functional) \cite{LB94}, the transformed potential no longer 
has the correct $(-1/r)$ asymptote \cite{Staroverov2b}. Further improvements on their scheme are needed to resolve this. 
Alternatively, one may adopt a density functional whose functional derivative has the correct $(-1/r)$ asymptote \cite{LFA}, which should be more desirable than AC model potentials. 
However, in this work, we only focus on the general performance of the AC model potential scheme. 

The rest of this paper is organized as follows. In Sec.\ II, we evaluate the performance of the LC hybrid scheme, AC model potential scheme, and other density functional methods, on a very wide range of applications, 
including the asymptote problems, SIE problems, energy-gap problems, charge-transfer problems, and many others. In Sec.\ III, we give our conclusions on the usefulness of these methods.

\section{Results and Discussion}

For a comprehensive comparison of various density functional methods, we examine the performance of the local spin density approximation (LSDA) \cite{LDA}, generalized gradient approximations (GGAs) 
(PBE \cite{PBE} and BLYP \cite{BLYP}), meta-generalized gradient approximations (MGGAs) (M06L \cite{M06L} and VS98 \cite{VS98}), 
global hybrid functionals (B3LYP \cite{B3LYP} and M06-2X \cite{M06-2X}), LC hybrid functionals ($\omega$B97 \cite{wB97X}, $\omega$B97X \cite{wB97X}, and $\omega$B97X-D \cite{wB97X-D}), 
and AC model potentials (LB94 \cite{LB94} and LB$\alpha$ \cite{LBa}), on various test sets involving 
the 223 atomization energies (AEs) of the G3/99 set \cite{G2a}, the 40 ionization potentials (IPs), 25 electron affinities (EAs), and 8 proton affinities (PAs) of the G2-1 set \cite{G21}, 
the 76 barrier heights (BHs) of the NHTBH38/04 and HTBH38/04 sets \cite{BH}, the 22 noncovalent interactions of the S22 set \cite{S22}, 
the additional 48 AEs of the G3/05 set \cite{G305}, the 113 AEs of the AE113 database \cite{Sup}, the 131 vertical IPs of the IP131 database \cite{wM05-D}, the 131 vertical EAs of the EA131 database \cite{wM05-D,Sup}, 
the 131 fundamental gaps (FGs) of the FG131 database \cite{wM05-D,Sup}, two dissociation curves of symmetric radical cations, 19 valence excitation energies, 23 Rydberg excitation energies, 
and one long-range charge-transfer excitation curve of two well-separated molecules. As discussed in Ref.\ \cite{wM05-D}, each vertical EA can be evaluated by two different ways, and each FG can be evaluated by 
three different ways, so there are in total 1386 pieces of data in our test sets, which are very large and diverse. Unspecified detailed information of the test sets, basis sets, and numerical grids used is 
given in Ref.\ \cite{wM05-D}. The LB94 (or LB$\alpha$) potential can be expressed as a linear combination of the LSDA exchange potential, the LSDA correlation potential, and a gradient-dependent exchange potential 
(e.g., see Eq.\ (55) of Ref.\ \cite{LB94}). Due to the inclusion of the gradient-dependent exchange potential, the LB94 (or LB$\alpha$) potential is {\it not} a functional derivative \cite{Staroverov,Staroverov2b}. 
In this work, the exchange energy from the LB94 (or LB$\alpha$) exchange potential is calculated by the popular Levy-Perdew virial relation \cite{virial}: 
\begin{equation}
E_{x}[\rho] = \int [3\rho(\textbf{r}) + \textbf{r} \cdot \nabla \rho(\textbf{r})] v_{x}(\textbf{r})d\textbf{r}, 
\end{equation} 
while the correlation energy from the LB94 (or LB$\alpha$) correlation potential is directly calculated by the LSDA correlation energy functional (i.e., its parent functional). 

All calculations are performed with a development version of \textsf{Q-Chem 3.2} \cite{QChem}. Spin-restricted theory is used for singlet states and spin-unrestricted theory for others, unless noted otherwise. 
For the binding energies of the weakly bound systems, the counterpoise correction \cite{CP} is employed to reduce basis set superposition error (BSSE). 
The error for each entry is defined as (error $=$ theoretical value $-$ reference value). 
The notation used for characterizing statistical errors is as follows: Mean signed errors (MSEs), mean absolute errors (MAEs), and root-mean-square (rms) errors.

\subsection{$\omega$B97 training set} 

The $\omega$B97 training set \cite{wB97X} consists of several popular databases, such as the 223 AEs of the G3/99 set \cite{G2a}, the 40 IPs, 25 EAs, and 8 PAs of the G2-1 set \cite{G21}, 
the 76 BHs of the NHTBH38/04 and HTBH38/04 sets \cite{BH}, and the 22 noncovalent interactions of the S22 set \cite{S22}. Table~\ref{table:training} summarizes the statistical errors of several methods 
for the $\omega$B97 training set. Results are computed using the 6-311++G(3df,3pd) basis set with the fine grid EML(75,302), consisting of 75 Euler-Maclaurin radial grid points \cite{EM} and 
302 Lebedev angular grid points \cite{L}. 

As shown, $\omega$B97, $\omega$B97X, and $\omega$B97X-D perform the best, as they all have very flexible functional forms that were parametrized using this training set. 
Although M06-2X was not particularly parametrized using this training set, it also performs very well. B3LYP performs similarly to M06L and VS98, and significantly outperforms LSDA, PBE, BLYP, LB94, and LB$\alpha$. 
Due to the lack of $E_{xc}[\rho]$, LB94 and LB$\alpha$ perform the worst (even worse than LSDA). 
Therefore, one should avoid using the AC model potential scheme for the calculation of total energies and related properties. 

\begin{table*}
	\caption{\label{table:training} Statistical errors (in kcal/mol) of the $\omega$B97 training set \cite{wB97X}.} 
	\begin{scriptsize}
	\begin{ruledtabular}
		\begin{tabular*}{\textwidth}{lcrrrrrrrrrrrr}
		System & Error & LSDA & PBE & BLYP & M06L & VS98 & B3LYP & M06-2X & $\omega$B97 & $\omega$B97X & $\omega$B97X-D & LB94 & LB$\alpha$\\ 
		\hline
		G3/99 & MSE & 120.59 & 20.90 & -4.59 & 2.33 &-1.11 &-4.32 & 0.36 & -0.29 & -0.20 & -0.24 & -482.58 & -56.44\\
		(223) & MAE & 120.60 & 21.51 & 9.76  & 5.37 & 3.46 & 5.49 & 2.30 & 2.63 & 2.13 & 1.93 & 484.91 & 103.84\\
		           & rms & 142.50 & 26.30 & 12.97 & 7.07 & 4.50 & 7.36 & 3.39 & 3.58 & 2.88 & 2.77 & 609.20 & 171.61\\
		\hline
		IP     & MSE & 3.42 & 0.03 &-1.51 &-1.88 & 0.52 & 2.17 & 0.16 & -0.51 & -0.15 & 0.19 & 97.67 & 104.03\\
		(40)  & MAE & 5.54 & 3.46 & 4.42 & 3.97 & 2.94 & 3.68 & 2.45 & 2.67 & 2.69 & 2.74 & 97.92 & 104.03\\
		          & rms & 6.66 & 4.35 & 5.27 & 4.95 & 3.66 & 4.79 & 3.58 & 3.60 & 3.59 & 3.61 & 116.03 & 115.94\\
		\hline
		EA	& MSE & 6.45 & 1.72 & 0.36 &-2.89 & 0.13 & 1.71 &-1.37 & -1.52 & -0.47 & 0.07 & 127.66 & 116.53\\
		(25)  & MAE & 6.45 & 2.42 & 2.57 & 3.83 & 2.61 & 2.38 & 2.56 & 2.72 & 2.04 & 1.91 & 127.66 & 116.53\\
		         & rms & 7.29 & 3.06 & 3.17 & 4.47 & 3.60 & 3.27 & 3.07 & 3.11 & 2.57 & 2.38 & 136.16 & 120.87\\
		\hline
		PA  & MSE &-5.91 &-0.83 &-1.47 & 0.97 & 1.00 &-0.77 &-1.21 & 0.67 & 0.56 & 1.42 & -122.75 & -95.66\\
		(8)  & MAE & 5.91 & 1.60 & 1.58 & 1.94 & 1.60 & 1.16 &2.02  & 1.48 & 1.21 & 1.50 & 122.75 & 95.66\\
		       & rms & 6.40 & 1.91 & 2.10 & 2.42 & 2.18 & 1.36 &2.21  & 2.18 & 1.70 & 2.05 & 126.54 & 98.45\\
		\hline
		NHTBH & MSE &-12.41 &-8.52 &-8.69 &-3.22 &-4.80 &-4.57 &-0.00 & 1.32 & 0.55 & -0.45 & -58.93 & -40.86\\
		(38)        & MAE & 12.62 & 8.62 & 8.72 & 3.73 & 5.16 & 4.69 &1.41 & 2.32 & 1.75 & 1.51 & 93.94 & 68.95\\
		               & rms & 16.13 &10.61 &10.27 & 5.26 & 6.49 & 5.71 &1.91 & 2.82 & 2.08 & 2.00 & 127.21 & 90.80\\
		\hline
		HTBH & MSE &-17.90 &-9.67 &-7.84 &-4.54 &-5.25 &-4.48 &-0.80  & -0.66 & -1.55 & -2.57 & 38.65 & 12.68\\
		(38)     & MAE & 17.90 & 9.67 & 7.84 & 4.54 & 5.26 & 4.56 & 1.32 & 2.11 & 2.27 & 2.70 & 44.31 & 22.80\\
		            & rms & 18.92 &10.37 & 8.66 & 5.22 & 5.92 & 5.10 & 1.56  & 2.47 & 2.60 & 3.10 & 57.37 & 29.90\\
		\hline
		S22	& MSE &-1.97 & 2.77 & 5.05 & 0.91 &-7.18 & 3.95 & 0.27  & 0.15 & 0.52 & -0.08 & 51.70 & 36.16\\
		(22)  & MAE & 2.08 & 2.77 & 5.05 & 0.92 & 7.63 & 3.95 & 0.39  & 0.59 & 0.86 & 0.21 & 51.70 & 36.16\\
		         & rms & 3.18 & 3.89 & 6.31 & 1.06 &11.80 & 5.17 & 0.55  & 0.77 & 1.28 & 0.26 & 61.66 & 42.55\\
		\hline
		Total	 & MSE & 65.86 & 10.32 &-4.07 & 0.27 &-1.91 &-2.79 & 0.05 &-0.23 &-0.21 &-0.38 &-256.11 & -16.49\\
		(394) & MAE & 72.41 & 14.63 & 8.05 & 4.57 & 3.88 & 4.77 & 2.04 & 2.42 & 2.07 & 1.94 & 310.76 &  89.67\\
		           & rms &107.53 & 20.40 &10.88 & 6.12 & 5.38 & 6.39 & 3.02 & 3.27 & 2.77 & 2.73 & 463.27 & 141.88\\
		\end{tabular*}
	\end{ruledtabular}
	\end{scriptsize}
\end{table*}

\subsection{Atomization energies} 

\subsubsection{G3/05 set} 

Here, we examine the performance of these functionals on the additional 48 AEs of the G3/05 set \cite{G305}, using the G3LargeXP basis set \cite{G3LXP} and the extra fine grid EML(99,590). 
As can be seen in Table~\ref{table:g305}, $\omega$B97, $\omega$B97X, and $\omega$B97X-D remain the best in performance, even if they are not particularly parametrized using this set. 
On the other hand, due to the lack of $E_{xc}[\rho]$, LB94 and LB$\alpha$ remain the worst in performance. 

\begin{table*}
	\caption{\label{table:g305} Statistical errors (in kcal/mol) of the additional 48 atomization energies of the G3/05 set \cite{G305}.} 
	\begin{scriptsize}
	\begin{ruledtabular}
		\begin{tabular*}{\textwidth}{lcrrrrrrrrrrrr}
		System & Error & LSDA & PBE & BLYP & M06L & VS98 & B3LYP & M06-2X & $\omega$B97 & $\omega$B97X & $\omega$B97X-D & LB94 & LB$\alpha$\\
		\hline
		G3/05 & MSE & 78.69 & 16.80 & 1.01 & 5.23 & 1.48 &-3.57 &-3.54 & 1.28 & 0.76 & 0.24 & -427.33 & -129.60\\
		(48)     & MAE & 78.83 & 18.07 & 9.89 & 7.50 & 4.62 & 6.29 & 5.14 & 4.25 & 3.60 & 3.01 & 439.45 & 155.28\\
		            & rms &115.99 & 30.22 &15.47 &10.05 & 6.62 & 8.90 & 6.66 & 5.41 & 4.52 & 3.95 & 863.35 & 395.46\\
		\end{tabular*}
	\end{ruledtabular}
	\end{scriptsize}
\end{table*}

\subsubsection{AE113 database} 

Recently, we developed the IP131, EA115, and FG115 databases, consisting of accurate reference values for the 131 vertical IPs, 115 vertical EAs, and 115 FGs, respectively, of 18 atoms and various molecules in 
their experimental geometries \cite{wM05-D}. To be consistent with the molecules collected in IP131, EA115 is expanded to EA131 (131 vertical electron affinities) \cite{wM05-D,Sup} to include the vertical EAs of the 
additional 16 molecules, and FG115 is expanded to FG131 (131 fundamental gaps) \cite{wM05-D,Sup} to include the FGs of the additional 16 molecules. In addition, we develop a new database 
AE113 (113 atomization energies) \cite{Sup} for the AEs of the 113 molecules in IP131. 

As described in Ref.\ \cite{wM05-D}, the reference values in EA131, FG131, and AE113 are obtained via the very accurate CCSD(T) theory (coupled-cluster theory with iterative singles and doubles and perturbative 
treatment of triple substitutions) \cite{CCSDT}, where the CCSD(T) correlation energies in the basis-set limit are extrapolated from calculations using the aug-cc-pVTZ and aug-cc-pVQZ basis sets \cite{extrap}: 
\begin{equation}
\label{eq:extrapolation}
E_{XY}^{\infty} = \frac{E_{X}^{\text{corr}} {X^3} - E_{Y}^{\text{corr}} {Y^3}}{{X^3}-{Y^3}}, 
\end{equation}
with $X = 3$ and $Y = 4$ for the aug-cc-pVTZ and aug-cc-pVQZ basis sets, respectively. Note that the reference atomization energies for the AE113 database are calculated without the zero-point energy correction. 
For completeness of this work, the reference values of IP131, EA131, FG131, and AE113 are all given in the supplementary material \cite{Sup}. 

The performance of the functionals is examined on the AE113 database, using the 6-311++G(3df,3pd) basis set and the fine grid EML(75,302). 
As shown in Table~\ref{table:AE113}, $\omega$B97, $\omega$B97X, and $\omega$B97X-D have the best performance, while LB94 and LB$\alpha$ have the worst performance, 
which is consistent with their performance on the atomization energies of the G3/99 and G3/05 sets. 

\begin{table*}
	\caption{\label{table:AE113} Statistical errors (in eV) of the atomization energies for the AE113 database.} 
	\begin{scriptsize}
	\begin{ruledtabular}
		\begin{tabular*}{\textwidth}{lcrrrrrrrrrrrr}
		System & Error & LSDA & PBE & BLYP & M06L & VS98 & B3LYP & M06-2X & $\omega$B97 & $\omega$B97X & $\omega$B97X-D & LB94 & LB$\alpha$\\
		\hline
		AE113 & MSE & 3.75 & 0.84 & 0.13 & 0.17 & 0.07 & -0.04 & 0.07 & 0.05 & 0.05 & 0.04 & -15.24 & -2.36\\
		(113)    & MAE & 3.75 & 0.88 & 0.36 & 0.27 & 0.14 & 0.16 & 0.12 & 0.12 & 0.10 & 0.10 & 15.53 & 4.81\\
		              & rms & 4.22 & 1.06 & 0.48 & 0.36 & 0.20 & 0.23 & 0.16 & 0.15 & 0.13 & 0.14 & 20.02 & 7.44\\
		\end{tabular*}
	\end{ruledtabular}
	\end{scriptsize}
\end{table*}

\subsection{Frontier orbital energies} 

For a system of $N$ electrons ($N$ is an integer), the vertical IP is defined as $\text{IP}(N) \equiv E(N-1) - E(N)$, and the vertical EA is defined as $\text{EA}(N) \equiv E(N) - E(N+1)$, with ${E}({N})$ being 
the ground-state energy of the $N$-electron system. For the exact KS-DFT, $\text{IP}(N)$ is identical to the minus HOMO energy of the $N$-electron system \cite{Janak,Fractional,Levy84,1overR,HOMO,HOMO2}, 
\begin{equation}\label{eqIP}
\text{IP}(N) = - {\epsilon}_{N}(N), 
\end{equation}
and $\text{EA}(N)$, which is IP($N+1$) by its definition, is identical to the minus HOMO energy of the ($N$+1)-electron system, 
\begin{equation}\label{eqEA}
\text{EA}(N) = - {\epsilon}_{N+1}(N+1), 
\end{equation} 
where ${\epsilon}_{i}(N)$ is the $i$-th KS orbital energy of the $N$-electron system. 

Conventionally, $\text{EA}(N)$ is approximated by the minus LUMO (lowest unoccupied molecular orbital) energy of the $N$-electron system, 
\begin{equation}\label{eqEA2}
\text{EA}(N) \approx - {\epsilon}_{N+1}(N). 
\end{equation} 
However, it has been found that there exists a difference between ${\epsilon}_{N+1}(N+1)$ and ${\epsilon}_{N+1}(N)$, 
\begin{equation}\label{eqDD2}
{\Delta}_{xc} = {\epsilon}_{N+1}(N+1) - {\epsilon}_{N+1}(N), 
\end{equation} 
arising from the derivative discontinuity (DD) of $E_{xc}[\rho]$ \cite{AC8,Fractional,DD1,DD2,DD2a,DD3,HOMO2,DD4a,DD0,DD6,DD8,DDChai}, 
\begin{equation}\label{eq:DD}
{\Delta}_{xc} = \lim_{\eta \to 0^{+}} \bigg\lbrace \frac{\delta E_{xc}[\rho]}{\delta \rho({\bf r})}\bigg |_{N+\eta} - \frac{\delta E_{xc}[\rho]}{\delta \rho({\bf r})}\bigg |_{N-\eta} \bigg\rbrace. 
\end{equation}

By contrast, as a hybrid functional, containing a fraction of the nonlocal HF exchange, belongs to the generalized KS (GKS) method, the corresponding GKS orbital energies incorporate part of 
the DD \cite{GKS1,DD7,Correction3,DDHirao,GKS2}. Recent study shows that ${\Delta}_{xc}$ (see Eq.\ (\ref{eqDD2})) is close to zero for LC hybrid functionals, so the $-{\epsilon}_{N+1}(N)$ calculated by 
LC hybrid functionals should be close to $\text{EA}(N)$ \cite{DDHirao}. 

Here, we evaluate the performance of the functionals on frontier orbital energies for the IP131 and EA131 databases using the 6-311++G(3df,3pd) basis set with the fine grid EML(75,302), and summarize 
the results in Table~\ref{table:frontier}. 

For the IP131 database, $\text{IP}(N)$ is calculated by $-{\epsilon}_{N}(N)$. As shown, the LSDA, GGAs, and MGGAs severely underestimate the vertical IPs, due to the incorrect asymptotic behavior 
of the XC potential. By contrast, the LC hybrid functionals and AC model potentials yield very accurate vertical IPs, due to the correct asymptote of the XC potential. 

For the EA131 database, $\text{EA}(N)$ is calculated by both $-{\epsilon}_{N+1}(N+1)$ and $-{\epsilon}_{N+1}(N)$. For the calculations using $-{\epsilon}_{N+1}(N+1)$ (the minus HOMO energy 
of the ($N$+1)-electron system), the LSDA, GGAs, and MGGAs seriously underestimate the vertical EAs, due to the incorrect asymptote of the XC potential, while the LC hybrid functionals and AC model potentials 
perform very well, due to the correct asymptote of the XC potential. 

On the other hand, for the calculations using $-{\epsilon}_{N+1}(N)$ (the minus LUMO energy of the $N$-electron system), the situation becomes very different. 
Although the AC model potentials yield accurate ${\epsilon}_{N+1}(N)$ due to the correct asymptote, the vertical EA remains severely overestimated due to the lack of the DD. 
In spite of the lack of the DD, there is a fortuitous cancellation of errors in the vertical EA predicted by the LSDA, GGAs, and MGGAs, as the ${\epsilon}_{N+1}(N)$ is (incorrectly) upshifted due to the incorrect asymptote 
of the XC potential, effectively incorporating part of the DD. 
As the DD is close to zero for LC hybrid functionals \cite{DDHirao}, i.e., ${\epsilon}_{N+1}(N+1) \approx {\epsilon}_{N+1}(N)$, the LC hybrid functionals significantly outperform the others using this estimate. 

\begin{table*}
	\caption{\label{table:frontier} Statistical errors (in eV) of the frontier orbital energies for the IP131 \cite{wM05-D} and EA131 databases.} 
	\begin{scriptsize}
	\begin{ruledtabular}
		\begin{tabular*}{\textwidth}{lcrrrrrrrrrrrr}
		System & Error & LSDA & PBE & BLYP & M06L & VS98 & B3LYP & M06-2X & $\omega$B97 & $\omega$B97X & $\omega$B97X-D & LB94 & LB$\alpha$\\
		\hline		
		\multicolumn{14}{c}{$-{\epsilon}_{N}(N) - \text{IP}_{\text{vertical}}$}\\
		Atoms & MSE & -4.88 & -4.87 & -5.00 & -4.66 & -4.80 & -3.71 & -2.22 & -0.99 & -1.17 & -1.63 & -0.21 & -0.72\\
		(18)     & MAE & 4.88 & 4.87 & 5.00 & 4.66 & 4.80 & 3.71 & 2.22 & 0.99 & 1.17 & 1.63 & 0.48 & 0.73\\
		             & rms & 5.23 & 5.22 & 5.31 & 4.94 & 5.12 & 3.94 & 2.32 & 1.37 & 1.52 & 1.97 & 0.63 & 0.95\\
		Molecules	 & MSE & -4.18 & -4.32 & -4.42 & -4.09 & -4.21 & -3.06 & -1.36 & -0.12 & -0.37 & -0.91 & 1.08 & 0.53\\
		(113)          & MAE & 4.18 & 4.32 & 4.42 & 4.09 & 4.21 & 3.06 & 1.36 & 0.31 & 0.41 & 0.91 & 1.09 &  0.58\\
		                    & rms & 4.24 & 4.38 & 4.47 & 4.14 & 4.26 & 3.10 & 1.40 & 0.41 & 0.53 & 1.00 & 1.18 & 0.68\\
		Total	 & MSE & -4.18 & -4.32 & -4.42 & -4.09 & -4.21 & -3.06 & -1.36 & -0.12 & -0.37 & -0.91 & 1.08 & 0.53\\
		(131) & MAE & 4.18 & 4.32 & 4.42 & 4.09 & 4.21 & 3.06 & 1.36 & 0.31 & 0.41 & 0.91 & 1.09 & 0.58\\
		 	 & rms & 4.24 & 4.38 & 4.47 & 4.14 & 4.26 & 3.10 & 1.40 & 0.41 & 0.53 & 1.00 & 1.18 & 0.68\\
		  \hline	 
		  \multicolumn{14}{c}{$-{\epsilon}_{N+1}(N+1) - \text{EA}_{\text{vertical}}$}\\         
		           Atoms & MSE & -2.84 & -2.89 & -3.02 & -2.93 & -3.01 & -2.24 & -1.32 & -0.17 & -0.25 & -0.53 & 0.08 & -0.32 \\
                                 (18) & MAE & 2.84 & 2.89 & 3.02 & 2.93 & 3.01 & 2.24 & 1.32 & 0.39 & 0.39 & 0.56 & 0.40 & 0.35 \\
                                         & rms & 3.05 & 3.10 & 3.22 & 3.14 & 3.21 & 2.40 & 1.42 & 0.65 & 0.66 & 0.83 & 0.59 & 0.56 \\
                      Molecules & MSE & -1.91 & -1.91 & -2.03 & -2.18 & -1.89 & -1.49 & -0.88 & -0.17 & -0.20 & -0.29 & 0.85 & 0.48 \\
                               (113) & MAE & 1.91 & 1.92 & 2.03 & 2.18 & 1.89 & 1.49 & 0.90 & 0.36 & 0.34 & 0.39 & 0.90 & 0.58 \\
                                          & rms & 2.15 & 2.17 & 2.25 & 2.37 & 2.17 & 1.64 & 0.95 & 0.42 & 0.43 & 0.49 & 1.07 & 0.77 \\
                                Total & MSE & -1.91 & -1.91 & -2.03 & -2.18 & -1.89 & -1.49 & -0.88 & -0.17 & -0.20 & -0.29 & 0.85 & 0.48 \\
                               (131) & MAE & 1.91 & 1.92 & 2.03 & 2.18 & 1.89 & 1.49 & 0.90 & 0.36 & 0.34 & 0.39 & 0.90 & 0.58 \\
                                          & rms & 2.15 & 2.17 & 2.25 & 2.37 & 2.17 & 1.64 & 0.95 & 0.42 & 0.43 & 0.49 & 1.07 & 0.77\\
		\hline		
		\multicolumn{14}{c}{$-{\epsilon}_{N+1}(N) - \text{EA}_{\text{vertical}}$}\\
		 Atoms & MSE & 3.11 & 2.62 & 2.77 & 2.02 & 2.47 & 2.08 & 0.77 & -0.53 & -0.36 & -0.02 & 8.00 & 7.36\\
		(18)     & MAE & 3.11 & 2.76 & 2.77 & 2.14 & 2.62 & 2.08 & 1.00 & 0.64 & 0.64 & 0.75 & 8.00 & 7.36\\
		             & rms & 3.54 & 3.16 & 3.19 & 2.54 & 3.02 & 2.36 & 1.14 & 0.85 & 0.85 & 0.89 & 8.58 & 7.85\\
		Molecules	 & MSE & 2.59 & 2.40 & 2.39 & 2.07 & 2.16 & 1.82 & 1.02 & -0.36 & -0.30 & 0.01 & 7.76 & 7.16\\
		(113)          & MAE & 2.59 & 2.40 & 2.39 & 2.07 & 2.16 & 1.82 & 1.02 & 0.47 & 0.50 & 0.51 & 7.76 & 7.16\\
		                    & rms & 2.87 & 2.65 & 2.62 & 2.34 & 2.46 & 2.02 & 1.10 & 0.54 & 0.58 & 0.58 & 7.98 & 7.37\\
		Total & MSE & 2.59 & 2.40 & 2.39 & 2.07 & 2.16 & 1.82 & 1.02 & -0.36 & -0.30 & 0.01 & 7.76 & 7.16\\
		(131) & MAE & 2.59 & 2.40 & 2.39 & 2.07 & 2.16 & 1.82 & 1.02 & 0.47 & 0.50 & 0.51 & 7.76 & 7.16\\
		           & rms & 2.87 & 2.65 & 2.62 & 2.34 & 2.46 & 2.02 & 1.10 & 0.54 & 0.58 & 0.58 & 7.98 & 7.37\\
		\end{tabular*}
	\end{ruledtabular}
	\end{scriptsize}
\end{table*}

\subsection{Fundamental gaps}

The FG of a $N$-electron system is defined as $E_{g} \equiv \text{IP}(N) - \text{EA}(N)$. As detailedly described in Ref.\ \cite{wM05-D}, since $\text{IP}(N)$ and $\text{EA}(N)$ can both be calculated by their definitions, or 
by their relations to the frontier orbital energies, $E_{g}$ can be evaluated by ${\epsilon}_{N+1}(N) - {\epsilon}_{N}(N)$, ${\epsilon}_{N+1}(N+1) - {\epsilon}_{N}(N)$, and $E(N-1) - 2 E(N) + E(N+1)$. 
Here, we examine the performance of the functionals on FGs for the FG131 database, using the 6-311++G(3df,3pd) basis set with the EML(75,302) grid, and summarize the results in Table~\ref{table:fundamental_gap}. 

For the calculations using ${\epsilon}_{N+1}(N) - {\epsilon}_{N}(N)$ (the HOMO-LUMO gaps), the LSDA, GGAs, MGGAs, and AC model potentials perform very poorly, due to the lack of the DD. 
As mentioned previously, the hybrid functionals incorporate part of the DD in the GKS orbital energies \cite{GKS1,GKS2,DD7,Correction3,DDHirao}, and therefore provide significant improvements, especially for 
the LC hybrid functionals. 

For the calculations using ${\epsilon}_{N+1}(N+1) - {\epsilon}_{N}(N)$ (the energy difference between the HOMOs of the $N$- and ($N+$1)-electron systems), the LC hybrid functionals and AC model potentials 
perform very well, due to the correct asymptote of the XC potential. In spite of the incorrect asymptote, there is a fortuitous cancellation of errors in the FGs calculated by the LSDA, GGAs, and MGGAs, as the HOMO energies 
of the $N$- and ($N+$1)-electron systems are both (incorrectly) upshifted. 

For the calculations using $E(N-1) - 2 E(N) + E(N+1)$ (the $\text{IP}(N) - \text{EA}(N)$ values), the AC model potentials perform very poorly, as the ground-state energies of the $N$- and ($N\pm$1)-electron systems 
are involved in this estimate. All the other functionals perform very well, as the ground-state energies are insensitive to the asymptote. 

\begin{table*}
	\caption{\label{table:fundamental_gap} Statistical errors (in eV) of the fundamental gaps for the FG131 database.} 
	\begin{scriptsize}
	\begin{ruledtabular}
		\begin{tabular*}{\textwidth}{lcrrrrrrrrrrrr}
		System & Error & LSDA & PBE & BLYP & M06L & VS98 & B3LYP & M06-2X & $\omega$B97 & $\omega$B97X & $\omega$B97X-D & LB94 & LB$\alpha$\\
		\hline
		\multicolumn{14}{c}{HOMO-LUMO gaps}\\
		Atoms & MSE & -7.93 & -7.43 & -7.70 & -6.61 & -7.20 & -5.73 &  -2.93 & -0.39 & -0.74 & -1.54 & -8.15 & -8.02\\
		(18)     & MAE & 7.93 & 7.43 & 7.70 & 6.61 & 7.20 & 5.73 & 2.93 & 0.77 & 1.04 & 1.78 & 8.15 & 8.02\\
		            & rms & 8.40 & 7.95 & 8.17 & 7.05 & 7.67 & 6.03 & 3.16 & 1.04 & 1.29 & 2.05 & 8.62 & 8.42\\
		Molecules	 & MSE & -6.91 & -6.86 & -6.95 & -6.29 & -6.51 & -5.02 &  -2.52 & 0.11 & -0.20 & -1.05 & -6.81 & -6.77\\
		(113)          & MAE & 6.91 & 6.86 & 6.95 & 6.29 & 6.51 & 5.02 & 2.52 & 0.45 & 0.49 & 1.06 & 6.81 & 6.77\\
		                   & rms & 7.08 & 7.01 & 7.09 & 6.44 & 6.67 & 5.12 & 2.57 & 0.57 & 0.65 & 1.28 & 7.06 & 6.99\\
		 Total & MSE &-6.91 & -6.86 & -6.95 & -6.29 & -6.51 & -5.02 & -2.52 & 0.11 & -0.20 & -1.05 & -6.81 & -6.77\\
		(131) & MAE & 6.91 & 6.86 & 6.95 & 6.29 & 6.51 & 5.02 & 2.52 & 0.45 & 0.49 & 1.06 & 6.81 & 6.77\\
		          & rms & 7.08 & 7.01 & 7.09 & 6.44 & 6.67 & 5.12 & 2.57 & 0.57 & 0.65 & 1.28 & 7.06 & 6.99\\
		\hline		           		
		\multicolumn{14}{c}{${\epsilon}_{N+1}(N+1) - {\epsilon}_{N}(N)$}\\
		Atoms & MSE & -1.98 & -1.92 & -1.91 & -1.66 & -1.73 & -1.40 & -0.83 & -0.75 & -0.85 & -1.04 & -0.22 & -0.33\\
		(18)     & MAE & 1.98 & 1.92 & 1.91 & 1.66 & 1.73 & 1.40 & 0.87 & 0.80 & 0.90 & 1.08 & 0.53 & 0.51\\
		             & rms & 2.36 & 2.33 & 2.28 & 1.98 & 2.10 & 1.66 & 1.00 & 1.05 & 1.12 & 1.30 & 0.73 & 0.67\\
		Molecules & MSE & -2.41 & -2.54 & -2.53 & -2.05 & -2.46 & -1.71 & -0.62 & -0.08 & -0.31 & -0.76 & 0.09 & -0.09\\
	          (113)         & MAE & 2.41 & 2.54 & 2.53 & 2.05 & 2.46 & 1.71 & 0.63 & 0.39 & 0.44 & 0.79 & 0.49 & 0.42\\
		                    & rms & 2.60 & 2.73 & 2.69 & 2.24 & 2.67 & 1.84 & 0.75 & 0.49 & 0.56 & 0.90 & 0.68 & 0.59\\
		Total  & MSE & -2.41 & -2.54 & -2.53 & -2.05 & -2.46 & -1.71 & -0.62 & -0.08 & -0.31 & -0.76 & 0.09 & -0.09\\
		(131) & MAE & 2.41 & 2.54 & 2.53 & 2.05 & 2.46 & 1.71 & 0.63 & 0.39 & 0.44 & 0.79 & 0.49 & 0.42\\
		           & rms & 2.60 & 2.73 & 2.69 & 2.24 & 2.67 & 1.84 & 0.75 & 0.49 & 0.56 & 0.90 & 0.68 & 0.59\\
                  \hline
		\multicolumn{14}{c}{$\text{IP}(N) - \text{EA}(N)$ values}\\
                   Atoms & MSE & 0.15 & 0.22 & 0.24 & 0.32 & 0.35 & 0.31 & 0.29 & 0.37 & 0.32 & 0.29 & 0.45 & 0.76 \\
		(18)     & MAE & 0.35 & 0.31 & 0.40 & 0.39 & 0.37 & 0.41 & 0.34 & 0.41 & 0.38 & 0.37 & 1.90 & 1.43\\
		             & rms & 0.58 & 0.51 & 0.60 & 0.62 & 0.64 & 0.60 & 0.51 & 0.66 & 0.61 & 0.59 & 2.67 & 2.16\\
		Molecules	  & MSE & -0.42 & -0.56 & -0.57 & -0.11 & -0.54 & -0.31 & 0.08 & 0.15 & 0.11 & -0.01 & -2.62  & -1.60 \\
	          (113)          & MAE & 0.52 & 0.62 & 0.62 & 0.38 & 0.61 & 0.40 & 0.29 & 0.33 & 0.31 & 0.29 & 2.86  & 1.90 \\
		                    & rms & 0.70 & 0.80 & 0.80 & 0.48 & 0.76 & 0.53 & 0.36 & 0.40 & 0.41 & 0.41 & 3.35  & 2.35 \\
		Total  & MSE & -0.42 & -0.56 & -0.57 & -0.11 & -0.54 & -0.31 & 0.08 & 0.15 & 0.11 & -0.01 & -2.62 & -1.60 \\
		(131) & MAE & 0.52 & 0.62 & 0.62 & 0.38 & 0.61 & 0.40 & 0.29 & 0.33 & 0.31 & 0.29 & 2.86 & 1.90 \\
		           & rms & 0.70 & 0.80 & 0.80 & 0.48 & 0.76 & 0.53 & 0.36 & 0.40 & 0.41 & 0.41 & 3.35 & 2.35 \\
		\end{tabular*}
	\end{ruledtabular}
	\end{scriptsize}
\end{table*}

\subsection{Dissociation of symmetric radical cations} 

Due to the severe SIEs of semilocal functionals, spurious fractional charge dissociation can occur \cite{SIE}, especially for symmetric charged radicals \ce{X_2^{+}}, such as \ce{H_2^{+}} \cite{h2p} 
and \ce{He_2^{+}} \cite{he2p}. To examine the performance of the density functional methods upon the SIE problems, we perform spin-unrestricted calculations using the aug-cc-pVQZ basis set and 
a high-quality EML(250,590) grid. The DFT results are compared with results from HF theory, and the highly accurate CCSD(T) theory. 

\begin{figure}
	\includegraphics[scale=0.6]{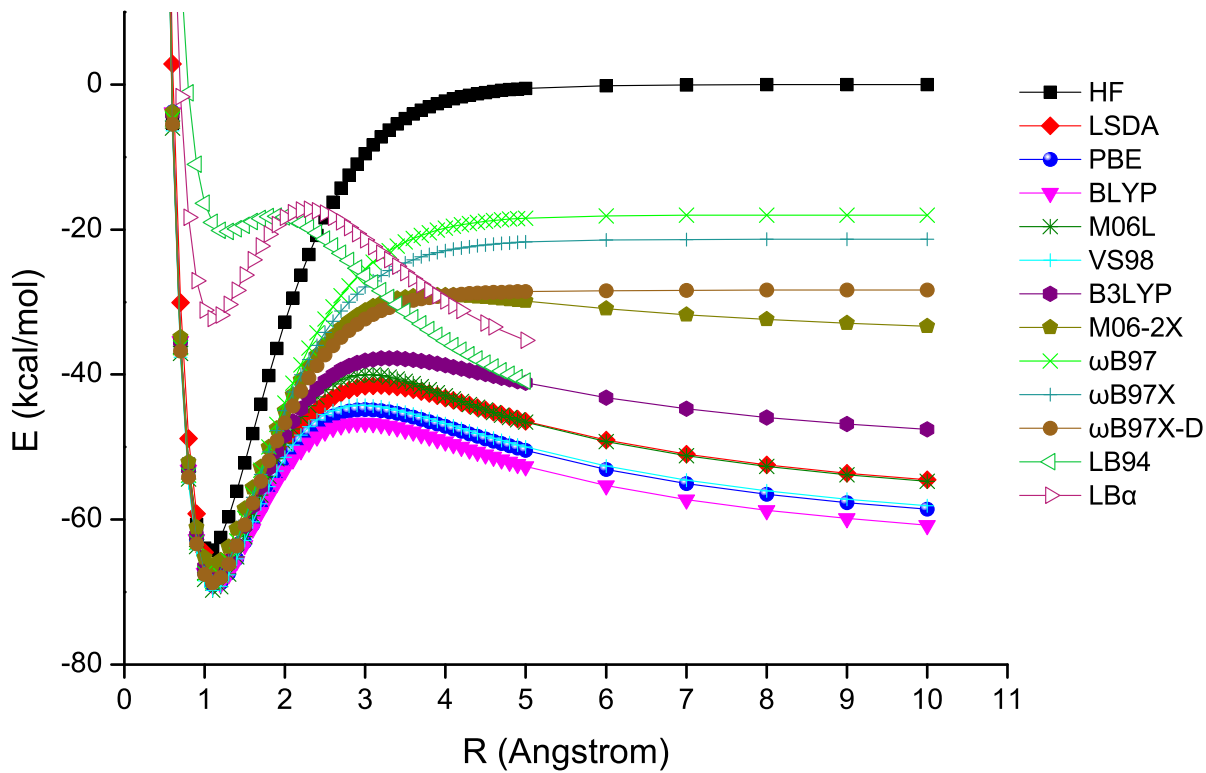}
	\caption{\label{fig:H} Dissociation curve of $\ce{H_{2}^{+}}$. Zero level is set to E(H)+E(H$^{+}$) for each method.}
\end{figure}

\begin{figure}
	\includegraphics[scale=0.6]{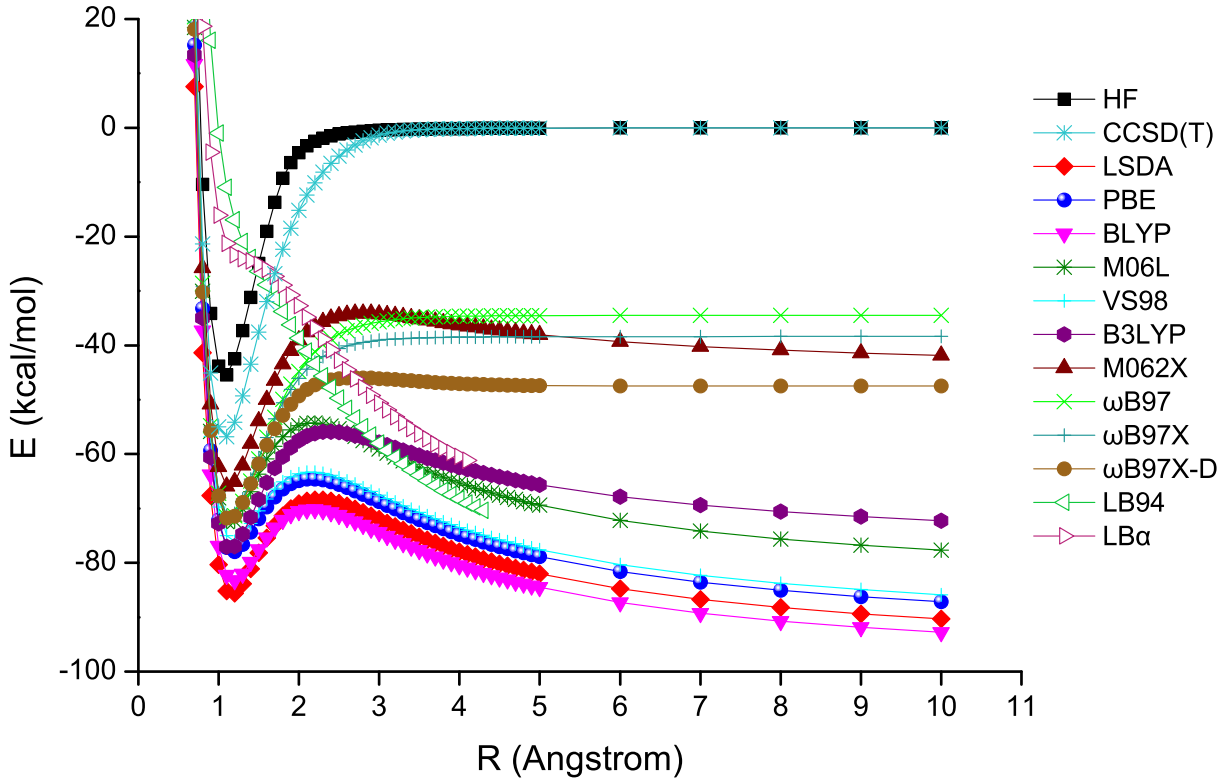}
	\caption{\label{fig:He} Dissociation curve of $\ce{He_{2}^{+}}$. Zero level is set to E(He)+E(He$^{+}$) for each method.} 
\end{figure}

The dissociation curves of \ce{H_2^{+}} and \ce{He_2^{+}} are shown in Figs.\ \ref{fig:H} and \ref{fig:He}, respectively. As can be seen, the LC hybrid scheme can remove the unphysical barriers 
of the dissociation curves, while the AC model potential scheme cannot. We emphasize that a SIE-corrected functional must be AC, while an AC functional (or model potential) may not be SIE-corrected. 

\subsection{Excitation energies} 

\subsubsection{Valence and Rydberg excitations}

To assess the performance of the density functionals on valence and Rydberg excitation energies, we perform TDDFT calculations on five small molecules \cite{Hirata}, including nitrogen gas ($\ce{N2}$), 
carbon monoxide ($\ce{CO}$), water ($\ce{H2O}$), ethylene ($\ce{C2H4}$), and formaldehyde ($\ce{CH2O}$), using the 6-311(2+,2+)G** basis set and EML(99,590) grid. The molecular geometries and 
experimental values of excitation energy are taken from Ref.\ \cite{Hirata}. For the AC model potential scheme, the adiabatic LSDA kernel is adopted in our TDDFT calculations \cite{LB94}. 

As shown in Table~\ref{table:TDDFT}, all the functionals perform reasonably well for the valence excitations. By contrast, for the Rydberg excitations, the LSDA and GGAs severely underestimate the excitation energies, 
and the MGGAs only provide very minor improvements. On the other hand, the LC hybrid functionals and AC model potentials yield very accurate excitation energies for Rydberg states, owing to the correct asymptote 
of the XC potential. 

\begin{table*}
	\begin{scriptsize}
	\caption{\label{table:TDDFT} The 19 valence and 23 Rydberg excitation energies (in eV) of \ce{N2}, \ce{CO}, water, ethylene, and formaldehyde, calculated by various methods. 
	The molecular geometries and experimental reference values are taken from Ref.\ \cite{Hirata}.} 
	\begin{ruledtabular}
		\begin{tabular*}{\textwidth}{llrrrrrrrrrrrrr}
		Molecule & State & LSDA & PBE & BLYP & M06L & VS98 & B3LYP & M06-2X & $\omega$B97 & $\omega$B97X & $\omega$B97X-D & LB94 & LB$\alpha$ & Expt.\\
		\hline
		\ce{N2}     & $\text{V}^{1}{\Pi}_\text{g}$ 			& 9.09 & 9.12 & 9.10 & 9.44 & 9.42 & 9.27 & 9.05 & 9.49 & 9.44 & 9.38 & 8.72 & 9.01 & 9.31\\
				& $\text{V}^{1}{\Sigma}_\text{u}^{-}$ 	& 9.70 & 9.70 & 9.60 & 9.85 & 10.03 & 9.35 & 8.43 & 9.36 & 9.30 & 9.31 & 9.46 & 9.53 & 9.97\\
				& $\text{V}^{1}{\Delta}_\text{u}$ 		& 10.27 & 10.12 & 9.90 & 10.29 & 10.16 & 9.75 & 10.01 & 9.85 & 9.82 & 9.82 & 10.06 & 10.13 & 10.27\\
				& $\text{V}^{3}{\Sigma}_\text{u}^{+}$ 	& 7.93 & 7.56 & 7.50 & 7.32 & 7.24 & 7.11 & 7.50 & 7.15 & 7.15 & 7.17 & 7.58 & 7.65 & 7.75\\
				& $\text{V}^{3}{\Pi}_\text{g}$ 			& 7.58 & 7.42 & 7.46 & 7.84 & 7.74 & 7.60 & 7.76 & 7.98 & 7.89 & 7.82 & 7.22 & 7.47 & 8.04\\
				& $\text{V}^{3}{\Delta}_\text{u}$ 		& 8.88 & 8.37 & 8.29 & 8.00 & 8.44 & 8.03 & 8.43 & 8.30 & 8.23 & 8.23 & 8.60 & 8.67 & 8.88\\
				& $\text{V}^{3}{\Sigma}_\text{u}^{-}$ 	& 9.70 & 9.70 & 9.60 & 10.29 & 10.03 & 9.35 & 8.90 & 9.36 & 9.30 & 9.31 & 9.46 & 9.53 & 9.67\\
				& $\text{V}^{3}{\Pi}_\text{u}$ 			& 10.36 & 10.40 & 10.32 & 11.07 & 10.80 & 10.64 & 11.36 & 11.14 & 11.05 & 10.98 & 10.13 & 10.26 & 11.19\\
		\hline
		\ce{CO}	& $\text{V}^{1}{\Pi}$				& 8.19 & 8.25 & 8.24 & 8.57 & 8.55 & 8.40 & 8.23 & 8.57 & 8.52 & 8.47 & 7.98 & 8.16 & 8.51\\
				& $\text{V}^{1}{\Sigma}^{-}$		& 9.89 & 9.86 & 9.78 & 10.26 & 10.19 & 9.72 & 9.12 & 9.88 & 9.80 & 9.78 & 9.92 & 9.94 & 9.88\\
				& $\text{V}^{3}{\Pi}$				& 5.96 & 5.74 & 5.82 & 6.10 & 6.03 & 5.86 & 6.29 & 6.15 & 6.11 & 6.07 & 5.60 & 5.75 & 6.32\\
				& $\text{V}^{3}{\Sigma}^{+}$		& 8.42 & 8.11 & 8.09 & 8.06 & 7.93 & 7.94 & 8.16 & 8.00 & 7.98 & 8.00 & 8.41 & 8.44 & 8.51\\
				& $\text{V}^{3}{\Delta}$			& 9.21 & 8.77 & 8.72 & 8.90 & 8.89 & 8.67 & 9.12 & 8.97 & 8.89 & 8.88 & 9.22 & 9.25 & 9.36\\
				& $\text{V}^{3}{\Sigma}^{-}$		& 9.89 & 9.86 & 9.78 & 10.51 & 10.19 & 9.72 & 9.31 & 9.88 & 9.80 & 9.78 & 9.92 & 9.94 & 9.88\\
		\hline
		\ce{H2O}  & $\text{R}^{1}\text{B}_{1}$ 	& 6.49 & 6.33 & 6.18 & 6.90 & 6.73 & 6.86 & 7.33 & 7.53 & 7.45 & 7.23 & 7.81 & 7.49 & 7.4\\
				& $\text{R}^{1}\text{A}_{2}$ 	& 7.62 & 7.44 & 7.26 & 7.70 & 7.78 & 8.23 & 8.81 & 9.19 & 9.09 & 8.63 & 9.66 & 9.28 & 9.1\\
				& $\text{R}^{1}\text{A}_{1}$ 	& 8.39 & 8.18 & 8.04 & 8.59 & 8.53 & 8.88 & 9.47 & 9.63 & 9.55 & 9.20 & 9.81 & 9.53 & 9.7\\
				& $\text{R}^{1}\text{B}_{1}$ 	& 7.99 & 7.84 & 7.66 & 8.19 & 8.09 & 8.75 & 9.58 & 9.81 & 9.69 & 9.17 & 11.14 & 10.53 & 10.0\\
				& $\text{R}^{1}\text{A}_{1}$ 	& 8.58 & 8.48 & 8.31 & 9.12 & 8.91 & 9.12 & 9.84 & 10.03 & 9.91 & 9.49 & 11.32 & 10.71 & 10.17\\
				& $\text{R}^{3}\text{B}_{1}$ 	& 6.24 & 6.00 & 5.89 & 6.60 & 6.38 & 6.49 & 6.98 & 7.12 & 7.09 & 6.89 & 7.33 & 7.06 & 7.2\\
		\hline
		\ce{C2H4}  	& $\text{R}^{1}\text{B}_\text{3u}$	& 6.61 & 6.40 & 6.16 & 6.63 & 6.63 & 6.58 & 6.88 & 7.57 & 7.41 & 7.02 & 7.77 & 7.38 & 7.11\\
					& $\text{V}^{1}\text{B}_\text{1u}$	& 7.41 & 7.29 & 7.08 & 7.46 & 7.45 & 7.35 & 7.51 & 7.63 & 7.60 & 7.52 & 7.70 & 7.66 & 7.60\\
					& $\text{R}^{1}\text{B}_\text{1g}$	& 7.08 & 6.87 & 6.61 & 7.01 & 7.06 & 7.10 & 7.39 & 8.19 & 8.05 & 7.59 & 7.22 & 7.53 & 7.80\\
					& $\text{R}^{1}\text{B}_\text{2g}$	& 7.05 & 6.83 & 6.55 & 7.00 & 7.03 & 7.09 & 7.47 & 8.33 & 8.16 & 7.66 & 8.46 & 8.03 & 8.01\\
					& $\text{R}^{1}\text{A}_\text{g}$ 	& 7.40 & 7.18 & 6.92 & 7.23 & 7.37 & 7.44 & 7.86 & 8.54 & 8.38 & 7.87 & 9.68 & 9.13 & 8.29\\
					& $\text{R}^{1}\text{B}_\text{3u}$	& 7.59 & 7.42 & 7.19 & 7.68 & 7.48 & 7.78 & 8.27 & 9.02 & 8.84 & 8.36 & 9.73 & 9.25 & 8.62\\
					& $\text{V}^{3}\text{B}_\text{1u}$        & 4.69 & 4.26 & 4.31 & 4.28 & 4.13 & 4.07 & 4.54 & 3.94 & 4.01 & 4.12 & 4.45 & 4.49 & 4.36\\
					& $\text{R}^{3}\text{B}_\text{3u}$       & 6.56 & 6.31 & 6.10 & 6.41 & 6.52 & 6.51 & 6.83 & 7.44 & 7.31 & 6.92 & 7.61 & 7.24 & 6.98\\
					& $\text{R}^{3}\text{B}_\text{1g}$	& 7.00 & 6.83 & 6.59 & 6.88 & 7.01 & 7.07 & 7.38 & 7.75 & 7.69 & 7.50 & 6.74 & 7.05 & 7.79\\
					& $\text{R}^{3}\text{B}_\text{2g}$       & 7.03 & 6.78 & 6.52 & 6.84 & 6.97 & 7.04 & 7.41 & 8.19 & 8.04 & 7.56 & 8.24 & 7.93 & 7.79\\
					& $\text{R}^{3}\text{A}_\text{g}$	         & 7.36 & 7.06 & 6.86 & 6.74 & 7.19 & 7.36 & 7.66 & 8.23 & 8.07 & 7.63 & 9.44 & 8.93 & 8.15\\
		\hline
		\ce{CH2O}	& $\text{V}^{1}\text{A}_{2}$ 	& 3.62 & 3.74 & 3.76 & 4.22 & 3.99 & 3.85 & 3.58 & 3.91 & 3.88 & 3.88 & 3.49 & 3.65 & 4.07\\
					& $\text{R}^{1}\text{B}_{2}$ 	& 5.89 & 5.77 & 5.63 & 6.25 & 6.10 & 6.47 & 7.15 & 7.43 & 7.35 & 6.96 & 7.43 & 7.13 & 7.11\\
					& $\text{R}^{1}\text{B}_{2}$ 	& 6.67 & 6.54 & 6.40 & 6.85 & 6.90 & 7.25 & 7.85 & 8.15 & 8.08 & 7.66 & 8.61 & 8.37 & 7.97\\
					& $\text{R}^{1}\text{A}_{1}$ 	& 7.22 & 7.11 & 6.94 & 7.42 & 7.37 & 7.98 & 9.01 & 9.30 & 9.21 & 8.74 & 9.58 & 9.51 & 8.14\\
					& $\text{R}^{1}\text{A}_{2}$ 	& 6.81 & 6.69 & 6.54 & 6.96 & 7.01 & 7.47 & 8.13 & 8.41 & 8.34 & 7.84 & 9.03 & 9.07 & 8.37\\
					& $\text{R}^{1}\text{B}_{2}$ 	& 6.90 & 6.81 & 6.63 & 7.20 & 7.00 & 7.71 & 8.91 & 9.15 & 8.99 & 8.52 & 10.00 & 9.67 & 8.88\\
					& $\text{V}^{3}\text{A}_{2}$ 	& 3.01 & 3.00 & 3.06 & 3.56 & 3.29 & 3.13 & 3.05 & 3.26 & 3.22 & 3.21 & 2.86 & 3.03 & 3.50\\
					& $\text{V}^{3}\text{A}_{1}$ 	& 6.03 & 5.57 & 5.59 & 5.49 & 5.39 & 5.22 & 5.49 & 5.27 & 5.25 & 5.29 & 5.87 & 5.89 & 5.86\\
					& $\text{R}^{3}\text{B}_{2}$ 	& 5.84 & 5.63 & 5.55 & 5.96 & 5.92 & 6.36 & 7.04 & 7.24 & 7.19 & 6.81 & 7.14 & 6.91 & 6.83\\
					& $\text{R}^{3}\text{B}_{2}$ 	& 6.65 & 6.47 & 6.36 & 6.66 & 6.82 & 7.17 & 7.69 & 7.91 & 7.87 & 7.50 & 8.43 & 8.17 & 7.79\\
					& $\text{R}^{3}\text{A}_{1}$ 	& 6.53 & 6.35 & 6.22 & 6.47 & 6.60 & 7.16 & 7.86 & 8.07 & 8.00 & 7.56 & 8.47 & 8.17 & 7.96\\
		\hline
		MAE & Valence (19) & 0.24 & 0.32 & 0.36 & 0.29 & 0.29 & 0.42 & 0.41 & 0.28 & 0.31 & 0.32 & 0.36 & 0.27 &  \\
			& Rydberg (23) & 1.12 & 1.30 & 1.48 & 1.04 & 1.03 & 0.75 & 0.29 & 0.26 & 0.20 & 0.35 & 0.73 & 0.42 &  \\
		\end{tabular*}
	\end{ruledtabular}
	\end{scriptsize}
\end{table*}

\subsubsection{Long-range charge-transfer excitations} 

Semilocal functionals qualitatively fail to describe long-range charge-transfer (CT) excitations between a donor and an acceptor \cite{Dreuw,Dreuw2,Dreuw3,Tozer,Gritsenko,fxcd}. 
The correct CT excitation energy ${\omega}_{\text{CT}}(R)$ should have the following asymptote \cite{Dreuw}: 
\begin{equation}\label{eq:CT}
{\omega}_{\text{CT}}(R\rightarrow\infty)\approx \text{IP}_{\text{D}} - \text{EA}_{\text{A}} - 1/R, 
\end{equation}
where $\text{IP}_{\text{D}}$ is the IP of the donor, $\text{EA}_{\text{A}}$ is the EA of the acceptor, and $R$ is their separation. As the ($-1/R$) dependency, which is the consequence of Coulomb interaction, can be 
accurately described by the nonlocal HF exchange \cite{Dreuw}, LC hybrid functionals, containing the full long-range HF exchange, can yield accurate CT excitation energies \cite{wB97X,wB97X-D,wM05-D,LC-D3}. 

On the other hand, within the framework of TDDFT, Hellgren and Gross showed that the discontinuity of the XC kernel is essential for the accurate description of CT excitations \cite{fxcd}. They showed that 
the divergency of the discontinuity as $r \rightarrow \infty$ can produce a divergent kernel in the dissociation limit, which is essential for the asymptote of CT excitation energy (see Eq.\ (\ref{eq:CT})) \cite{Gritsenko,fxcd}. 

Following Dreuw {\it et al.}, we perform TDDFT calculations for the lowest CT excitations between ethylene and tetrafluoroethylene, when separated by a distance $R$. For the AC model potentials, we adopt 
the adiabatic LSDA kernel in the TDDFT calculations \cite{LB94}. Calculations are performed with the 6-31G* basis and EML(99,590) grid. For comparison, high-level results by the symmetry-adapted-cluster 
configuration-interaction (SAC-CI) method are taken from Tawada {\it et al.} \cite{Tawada}. 

\begin{figure}
	\includegraphics[scale=1.0]{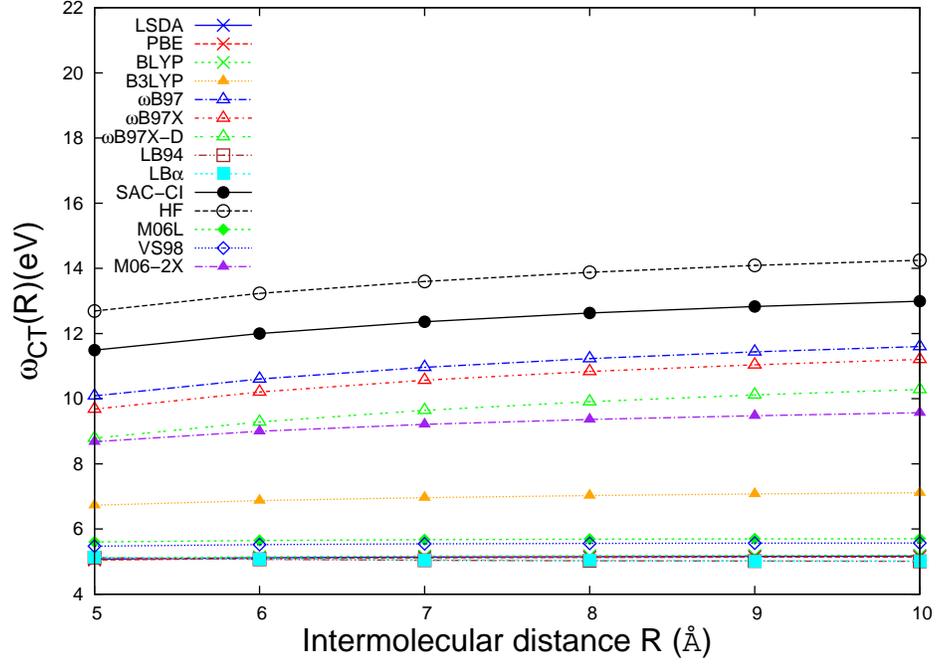}
	\caption{\label{fig:CTabsolute} The lowest CT excitation energy of \ce{C2H4}$\cdots$\ce{C2F4} dimer along the intermolecular distance $R$ (in $\text{\AA}$).} 
\end{figure}

Fig.\ \ref{fig:CTabsolute} shows the values of CT excitation energies. For the LSDA, GGAs, MGGAs, and AC model potentials, the predicted CT excitation energies are severely underestimated due to the lack of 
the discontinuity of the XC kernel. By incorporating a fraction of nonlocal HF exchange, the hybrid functionals show some improvements, especially for the LC hybrid functionals. 
However, since the LC hybrid functionals still have short-range SIEs, a more flexible operator for HF exchange may be needed in the LC hybrid functional to further reduce such errors \cite{op}. 

\begin{figure}
	\includegraphics[scale=1.0]{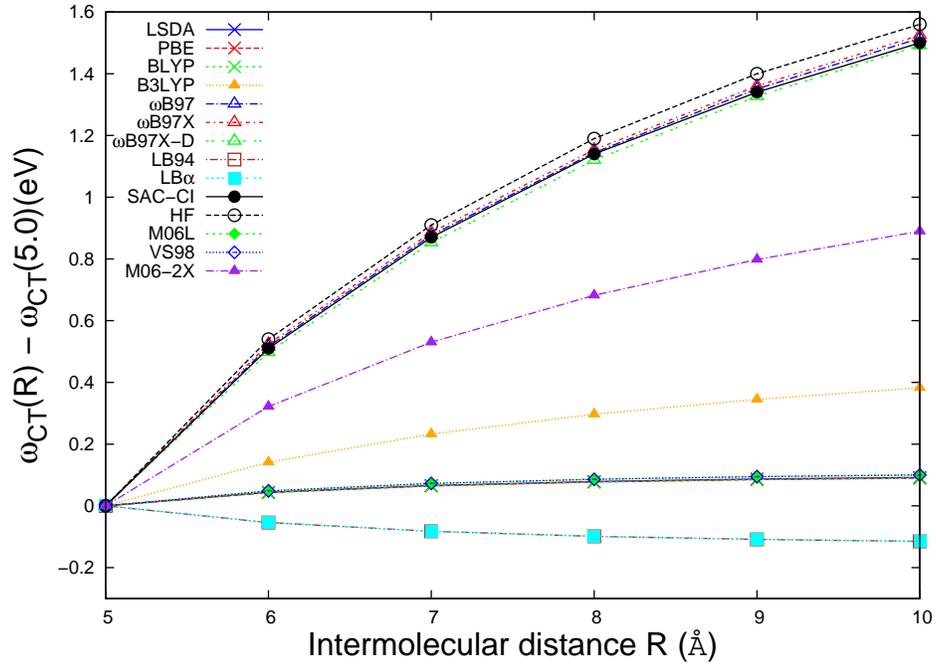}
	\caption{\label{fig:CTrelative} Relative excitation energy for the lowest CT excitation of \ce{C2H4}$\cdots$\ce{C2F4} dimer along the intermolecular distance $R$ (in $\text{\AA}$). 
	The excitation energy at 5 $\text{\AA}$ is set to zero for each method.} 
\end{figure}

Fig.\ \ref{fig:CTrelative} shows the relative values of the CT excitation energies, where the excitation energy at 5 $\text{\AA}$ is set to zero for each method. Therefore, the $(-1/R)$ dependence dominates 
the relative CT excitation energies. The results of the LSDA, GGAs, and MGGAs are overlapped, showing that none of them can capture the long-range Coulomb interaction in the CT excitations. 
The AC model potentials fail to improve semilocal functionals here. While the global hybrid functionals provide some improvements, the LC hybrid functionals predict the relative values of the CT excitation curves 
that are in an excellent agreement with the high-level SAC-CI results. 

\section{Conclusions} 

In conclusion, we have examined the performance of the LC hybrid scheme and AC model potential scheme on a very wide range of applications. 
Due to the correct XC potential in the asymptotic region, both the LC hybrid scheme and AC model potential scheme are reliably accurate for properties related to the asymptote, 
such as the HOMO energies and Rydberg excitation energies of molecules. 

Relative to the AC model potential scheme, the LC hybrid scheme has several advantages. 
First, the LC hybrid scheme shows significant improvements in the SIE problems, such as dissociation of symmetric radical cations, while the AC model potential scheme fails to resolve the SIE problems. 
Secondly, as the LC hybrid scheme belongs to the GKS method, the fundamental gap can be accurately predicted by the HOMO-LUMO gap. For the AC model potential scheme, the HOMO-LUMO gap 
is a poor approximation of the fundamental gap, due to the lack of the DD. 
Thirdly, the long-range HF exchange kernel in the LC hybrid scheme is essential for the accurate description of the long-range CT excitations between two well-separated molecules, which cannot be 
described by the AC model potential scheme due to the lack of the discontinuity of the XC kernel. 
Finally, the LC hybrid scheme yields accurate predictions for atomization energies, barrier heights, and noncovalent interactions. However, due to the lack of $E_{xc}[\rho]$, the AC model potential scheme 
performs very poorly for total energies and related properties. 

Despite their similarity in correct asymptotic behavior, the AC model potential scheme may be accurate only for the asymptote problems, while the LC hybrid scheme, which has remedied several 
qualitative failures of semilocal functionals, could be reliably accurate for a very wide range of applications. Although only three LC hybrid functionals ($\omega$B97, $\omega$B97X, and $\omega$B97X-D) are examined 
in this work, we expect that other LC hybrid functionals should also perform well for the properties sensitive to the long-range HF exchange, such as the asymptote problems, SIE problems, energy-gap problems, and 
charge-transfer problems. For the properties insensitive to the long-range HF exchange (e.g., atomization energies, noncovalent interactions), further comparisons should be made 
for other LC hybrid functionals (e.g., see Refs.\ \cite{wM05-D,LC-D3}).

\begin{acknowledgments}

This work was supported by the National Science Council of Taiwan (Grant No. NSC101-2112-M-002-017-MY3), National Taiwan University (Grant Nos.\ 99R70304, 101R891401, and 101R891403), 
and the National Center for Theoretical Sciences of Taiwan. Y.C.S. was partially supported by the NSC Undergraduate Fellowship. 
We are grateful for the partial support of computer resources from the groups of Dr. J.-L. Kuo (Academia Sinica) and Dr. Y.-C. Cheng (NTU). 

\end{acknowledgments}

\newpage

\section*{Supplementary material}

\begingroup
\squeezetable

\begin{ThreePartTable}

\begin{TableNotes}
\item [a] \label{tn:a} The reference values for the IP131 database, which were published in [Y.-S. Lin, C.-W. Tsai, G.-D. Li, and J.-D. Chai, \textit{J. Chem. Phys.}, 2012, {\bf 136}, 154109], are listed for completeness of this work. 
\item [b] \label{tn:*}   Most of the reference values for the EA131 and FG131 databases were published in the EA115 and FG115 databases [Y.-S. Lin, C.-W. Tsai, G.-D. Li, and J.-D. Chai, \textit{J. Chem. Phys.}, 
2012, {\bf 136}, 154109]. For clarity, the reference values for the additional 16 molecules are labelled with the ($^\text{b}$). 
\end{TableNotes}

\begin{longtable}{@{\hspace{20pt}}>{\raggedright}m{140pt}>{\centering}m{50pt}>{\centering}m{50pt}>{\centering}m{50pt}>{\centering\arraybackslash}m{50pt}}
\caption{\label{table:IP131_reference} Reference values (in eV) for the IP131\tnotex{tn:a}, EA131\tnotex{tn:*}, FG131\tnotex{tn:*}, and AE113 databases. For the CCSD(T) calculations, the correlation energies 
in the basis-set limit are extrapolated from calculations using the aug-cc-pVTZ and aug-cc-pVQZ basis sets. The reference atomization energies for AE113 are calculated without the zero-point energy correction. 
All the databases contain the 18 atoms and 113 molecules collected in IP131.}\\ 
\hline \hline
& IP131 & EA131 & FG131 & AE113\\
\hline
Molecule & Experimental Vertical Ionization Potential & CCSD(T) Vertical Electron Affinity & CCSD(T) Fundamental Gap & CCSD(T) Atomization Energy\\
\hline
\ce{H} (Hydrogen atom) & 13.60 & 0.75 & 12.86 & \\
\ce{He} (Helium atom) & 24.59 & -2.63 & 27.23 & \\
\ce{Li} (Lithium atom) & 5.39 & 0.62 & 4.22 & \\
\ce{Be} (Beryllium atom) & 9.32 & -0.36 & 9.66 & \\
\ce{B} (Boron atom) & 8.30 & 0.25 & 7.99 & \\
\end{longtable}

\end{ThreePartTable}

\endgroup
\pagebreak

\begingroup
\squeezetable
\begin{longtable*}{@{\hspace{10pt}}p{50pt}rrrrrrrrrrrrr}
\caption{\label{table:AE113} Atomization energies (in eV) of the AE113 database. The atomization energies are calculated without the zero-point energy correction.}\\
\hline \hline
Molecule & LSDA & PBE & BLYP & M06L & VS98 & B3LYP & M06-2X & $\omega$B97 & $\omega$B97X & $\omega$B97X-D & LB94 & LB$\alpha$ & Ref.\\
\hline
\ce{CH3} & 14.71 & 13.44 & 13.32 & 13.31 & 13.44 & 13.45 & 13.29 & 13.27 & 13.30 & 13.31 & 7.84 & 12.67 & 13.32 \\
\ce{CH4} & 20.04 & 18.21 & 18.08 & 18.14 & 18.24 & 18.26 & 18.15 & 18.11 & 18.14 & 18.16 & 9.69 & 16.58 & 18.19 \\
\ce{NH} & 4.15 & 3.84 & 3.89 & 3.59 & 3.60 & 3.82 & 3.61 & 3.66 & 3.65 & 3.65 & 4.81 & 5.13 & 3.60 \\
\ce{NH2} & 9.02 & 8.18 & 8.23 & 7.86 & 7.92 & 8.16 & 7.86 & 7.94 & 7.94 & 7.95 & 8.73 & 10.25 & 7.91 \\
\ce{NH3} & 14.63 & 13.10 & 13.09 & 12.66 & 12.82 & 13.06 & 12.84 & 12.86 & 12.87 & 12.90 & 12.22 & 15.54 & 12.91 \\
\ce{OH} & 5.40 & 4.78 & 4.77 & 4.55 & 4.67 & 4.71 & 4.63 & 4.60 & 4.63 & 4.65 & 5.86 & 6.58 & 4.65 \\
\hline \hline
\end{longtable*}
\endgroup

\newpage
\begingroup
\squeezetable
\begin{longtable*}{@{\hspace{10pt}}p{50pt}rrrrrrrrrrrrr}
\caption{\label{table:IP131_HOMO} $-{\epsilon}_{N}(N)$ (in eV) of the IP131 database.}\\
\hline \hline
Molecule & LSDA & PBE & BLYP & M06L & VS98 & B3LYP & M06-2X& $\omega$B97 & $\omega$B97X & $\omega$B97X-D & LB94 & LB$\alpha$ & Ref.\\
\hline
\ce{H} & -7.32 & -7.59 & -7.40 & -7.92 & -7.71 & -8.77 & -10.32 & -11.91 & -11.62 & -11.06 & -11.96 & -11.45 & 13.60 \\
\ce{He} & -15.52 & -15.76 & -15.91 & -16.53 & -16.16 & -18.00 & -20.62 & -21.13 & -20.90 & -20.26 & -23.15 & -22.22 & 24.59 \\
\ce{Li} & -3.16 & -3.22 & -3.03 & -3.11 & -3.20 & -3.65 & -4.18 & -5.32 & -5.31 & -5.21 & -5.24 & -5.07 & 5.39 \\
\ce{Be} & -5.60 & -5.61 & -5.47 & -5.71 & -5.66 & -6.32 & -7.26 & -8.76 & -8.63 & -8.29 & -8.71 & -8.40 & 9.32 \\
\ce{B} & -4.11 & -4.17 & -4.06 & -4.45 & -4.10 & -5.17 & -6.17 & -7.91 & -7.73 & -7.24 & -8.16 & -7.76 & 8.30 \\
\ce{C} & -6.13 & -6.10 & -5.95 & -6.36 & -6.20 & -7.34 & -8.84 & -10.36 & -10.14 & -9.57 & -10.92 & -10.50 & 11.26 \\
\hline \hline
\end{longtable*}
\endgroup

\newpage
\begingroup
\squeezetable
\begin{longtable*}{@{\hspace{10pt}}p{50pt}rrrrrrrrrrrrr}
\caption{\label{table:IP131_anion_homo} $-{\epsilon}_{N+1}(N+1)$ (in eV) of the EA131 database.}\\
\hline \hline
Molecule & LSDA & PBE & BLYP & M06L & VS98 & B3LYP & M06-2X& $\omega$B97 & $\omega$B97X & $\omega$B97X-D & LB94 & LB$\alpha$ & Ref.\\
\hline
\ce{H} & 1.87 & 1.97 & 2.05 & 1.64 & 1.98 & 1.21 & 0.32 & -1.37 & -1.21 & -0.85 & -0.80 & -0.42 & 0.75 \\
\ce{He} & 6.30 & 6.06 & 6.31 & 6.15 & 5.89 & 5.73 & 5.22 & 4.10 & 4.15 & 4.26 & 4.25 & 4.26 & -2.63 \\
\ce{Li} & 0.82 & 0.88 & 1.00 & 0.96 & 0.95 & 0.58 & 0.16 & -0.69 & -0.73 & -0.70 & -0.51 & -0.34 & 0.62 \\
\ce{Be} & 1.92 & 1.89 & 2.08 & 1.97 & 2.04 & 1.63 & 1.27 & 0.44 & 0.34 & 0.40 & 0.22 & 0.46 & -0.36 \\
\ce{B} & 2.08 & 2.14 & 2.32 & 2.20 & 2.33 & 1.60 & 0.77 & -0.45 & -0.42 & -0.15 & -0.42 & -0.10 & 0.25 \\
\ce{C} & 1.95 & 2.06 & 2.26 & 2.25 & 2.21 & 1.27 & 0.10 & -1.28 & -1.16 & -0.74 & -1.33 & -0.94 & 1.25 \\
\hline \hline
\end{longtable*}
\endgroup

\newpage
\begingroup
\squeezetable
\begin{longtable*}{@{\hspace{10pt}}p{50pt}rrrrrrrrrrrrr}
\caption{\label{table:IP131_LUMO} $-{\epsilon}_{N+1}(N)$ (in eV) of the EA131 database.}\\
\hline \hline
Molecule & LSDA & PBE & BLYP & M06L & VS98 & B3LYP & M06-2X& $\omega$B97 & $\omega$B97X & $\omega$B97X-D & LB94 & LB$\alpha$ & Ref.\\
\hline
\ce{H} & -2.60 & 0.51 & -0.85 & 0.41 & 0.53 & -1.12 & 0.52 & 0.52 & 0.53 & 0.53 & -2.48 & -2.45 & 0.75 \\
\ce{He} & 2.49 & 2.29 & 2.02 & 2.54 & 2.67 & 2.39 & 2.67 & 4.55 & 4.64 & 4.30 & -2.44 & -2.22 & -2.63 \\
\ce{Li} & -2.03 & -1.34 & -1.57 & -1.14 & -1.58 & -1.48 & -0.26 & 0.28 & 0.27 & 0.20 & -7.79 & -7.57 & 0.62 \\
\ce{Be} & -2.10 & -2.02 & -1.93 & -1.67 & -2.08 & -1.43 & -0.53 & 0.67 & 0.54 & 0.29 & -5.36 & -5.01 & -0.36 \\
\ce{B} & -3.92 & -3.62 & -3.45 & -2.84 & -3.56 & -2.69 & -1.40 & 0.15 & -0.12 & -0.57 & -8.32 & -7.73 & 0.25 \\
\ce{C} & -6.06 & -5.59 & -5.34 & -4.89 & -5.35 & -4.29 & -2.72 & -1.00 & -1.38 & -1.98 & -11.44 & -10.68 & 1.25 \\
\hline \hline
\end{longtable*}
\endgroup

\newpage
\begingroup
\squeezetable
\begin{longtable*}{@{\hspace{10pt}}p{50pt}rrrrrrrrrrrrr}
\caption{\label{table:IP131_HOMO_LUMO} HOMO-LUMO gaps (in eV) of the FG131 database.}\\
\hline \hline
Molecule & LSDA & PBE & BLYP & M06L & VS98 & B3LYP & M06-2X & $\omega$B97 & $\omega$B97X & $\omega$B97X-D & LB94 & LB$\alpha$ & Ref.\\
\hline
\ce{H} & 4.72 & 8.11 & 6.55 & 8.33 & 8.24 & 7.66 & 10.84 & 12.44 & 12.15 & 11.59 & 9.48 & 9.00 & 12.86 \\
\ce{He} & 18.01 & 18.05 & 17.93 & 19.07 & 18.83 & 20.39 & 23.30 & 25.68 & 25.54 & 24.57 & 20.71 & 20.00 & 27.23 \\
\ce{Li} & 1.13 & 1.88 & 1.46 & 1.97 & 1.61 & 2.17 & 3.93 & 5.60 & 5.57 & 5.41 & -2.55 & -2.50 & 4.22 \\
\ce{Be} & 3.50 & 3.59 & 3.54 & 4.03 & 3.58 & 4.89 & 6.72 & 9.43 & 9.17 & 8.58 & 3.35 & 3.40 & 9.66 \\
\ce{B} & 0.19 & 0.55 & 0.62 & 1.61 & 0.54 & 2.48 & 4.77 & 8.07 & 7.61 & 6.67 & -0.16 & 0.03 & 7.99 \\
\ce{C} & 0.08 & 0.52 & 0.61 & 1.47 & 0.86 & 3.05 & 6.12 & 9.37 & 8.75 & 7.59 & -0.52 & -0.18 & 9.97 \\
\hline \hline
\end{longtable*}
\endgroup

\newpage
\begingroup
\squeezetable
\begin{longtable*}{@{\hspace{10pt}}p{50pt}rrrrrrrrrrrrr}
\caption{\label{table:IP131_HOMO_HOMO} ${\epsilon}_{N+1}(N+1) - {\epsilon}_{N}(N)$ (in eV) of the FG131 database.}\\
\hline \hline
Molecule & LSDA & PBE & BLYP & M06L & VS98 & B3LYP & M06-2X & $\omega$B97 & $\omega$B97X & $\omega$B97X-D & LB94 & LB$\alpha$ & Ref.\\
\hline
\ce{H} & 9.19 & 9.56 & 9.46 & 9.56 & 9.69 & 9.99 & 10.64 & 10.54 & 10.41 & 10.21 & 11.16 & 11.03 & 12.86 \\
\ce{He} & 21.81 & 21.82 & 22.22 & 22.68 & 22.05 & 23.73 & 25.85 & 25.22 & 25.05 & 24.53 & 27.40 & 26.47 & 27.23 \\
\ce{Li} & 3.98 & 4.10 & 4.03 & 4.07 & 4.15 & 4.23 & 4.35 & 4.63 & 4.57 & 4.51 & 4.72 & 4.73 & 4.22 \\
\ce{Be} & 7.51 & 7.50 & 7.55 & 7.68 & 7.71 & 7.95 & 8.53 & 9.21 & 8.97 & 8.69 & 8.93 & 8.86 & 9.66 \\
\ce{B} & 6.19 & 6.31 & 6.38 & 6.65 & 6.44 & 6.77 & 6.94 & 7.46 & 7.32 & 7.09 & 7.74 & 7.66 & 7.99 \\
\ce{C} & 8.08 & 8.16 & 8.20 & 8.62 & 8.41 & 8.62 & 8.94 & 9.08 & 8.98 & 8.83 & 9.59 & 9.56 & 9.97 \\
\hline \hline
\end{longtable*}
\endgroup

\newpage
\begingroup
\squeezetable
\begin{longtable*}{@{\hspace{10pt}}p{50pt}rrrrrrrrrrrrr}
\caption{\label{table:IP131_IP_EA} $\text{IP}(N) - \text{EA}(N)$ values (in eV) of the FG131 database.}\\
\hline \hline
Molecule & LSDA & PBE & BLYP & M06L & VS98 & B3LYP & M06-2X & $\omega$B97 & $\omega$B97X & $\omega$B97X-D & LB94 & LB$\alpha$ & Ref.\\
\hline
\ce{H} & 12.15 & 12.95 & 12.76 & 13.10 & 13.03 & 12.80 & 12.98 & 13.00 & 12.92 & 12.90 & 10.96 & 11.48 & 12.86 \\
\ce{He} & 28.23 & 28.27 & 28.74 & 28.75 & 28.51 & 28.78 & 28.47 & 28.85 & 28.78 & 28.73 & 32.23 & 31.11 & 27.23 \\
\ce{Li} & 4.87 & 5.06 & 5.06 & 4.51 & 4.87 & 5.06 & 4.89 & 4.91 & 4.81 & 4.85 & 4.53 & 4.64 & 4.22 \\
\ce{Be} & 9.17 & 9.16 & 9.28 & 9.26 & 9.45 & 9.34 & 9.52 & 9.46 & 9.31 & 9.18 & 9.71 & 10.01 & 9.66 \\
\ce{B} & 7.97 & 8.13 & 8.27 & 8.35 & 8.06 & 8.33 & 8.10 & 8.30 & 8.29 & 8.23 & 9.07 & 8.78 & 7.99 \\
\ce{C} & 9.92 & 9.99 & 10.09 & 10.10 & 10.17 & 10.19 & 10.09 & 10.19 & 10.17 & 10.13 & 9.96 & 10.10 & 9.97 \\
\hline \hline
\end{longtable*}

\end{document}